\documentclass[11pt,a4paper]{article}

\usepackage{jheparxiv}
\usepackage{basket}
\usepackage{mathShrtcts}

\title{Chiral Splitting and $\mathcal N = 4$ Einstein--Yang--Mills Tree Amplitudes in 4d}
\author{Kai A. Roehrig}
\affiliation{Department of Applied Mathematics \& Theoretical Physics \\
University of Cambridge\\
Wilberforce Road\\
Cambridge CB3 0WA, United Kingdom}
\date{\today}

\abstract{We present a world-sheet formula for all tree level scattering amplitudes, in all trace sectors, of four dimensional $\mathcal{N} \leq 4$ supersymmetric Einstein-Yang-Mills theory, based on the refined scattering equations. This generalizes previously known formulas for all-trace purely bosonic, or supersymmetric single-trace amplitudes. We find this formula by applying a new chiral splitting formula for all CHY Pfaffians in 4d, into two determinants, of positive and negative helicity respectively. The splitting of CHY Pfaffians is shown to be a special case of the splitting of $T\mathbb{M}$ valued fermion correlators on the sphere, which does not require the scattering equations to hold, and is a consequence of the isomorphism $T\mathbb{M} \simeq \mathbb{S}^+ \otimes \mathbb{S}^-$ between the tangent bundle of Minkowski space and the left- and right-handed spin bundles. We present and prove this general splitting formula.}

\begin{document}
 \maketitle
 
 \section{Introduction}

 There have been fascinating new developments in the understanding of perturbative scattering amplitudes which have at their heart a map from the space of kinematics of $n$ massless particles to the moduli space of $n$-punctured Riemann surfaces. This map is provided by the scattering equations, and it allows the reformulation of tree and loop level scattering amplitudes of several quantum field theories in terms of integrals over said moduli space.

 These new representations come in two very different flavours: On the one hand there are intrinsically $4d$ representations like the Roiban-Spradlin-Volovich-Witten formula \cite{Witten:2003nn,Roiban:2004yf} for \Neqfour super-Yang-Mills and the Cachazo-Skinner formula for \Neqeight supergravity \citep{Cachazo:2012kg}. They have their origin in twistor theory, use spinor-helicity variables and accommodate supersymmetry rather naturally via the use of on-shell supersymmetry. On the other hand is the dimension agnostic Cachazo-He-Yuan framework, which has its roots in ambi-twistor space and can describe a plethora of scattering amplitudes in any number of space-time dimensions, but so far is largely limited to bosonic states.
 
It is a rather non-trivial fact that the CHY formulae reduce to the corresponding twistor formulae once the external kinematics is four dimensional. Both are underpinned by the same set of equations, albeit in very different representations, which is widely understood \cite{Roiban:2004yf,Witten:2004cp,Cachazo:2013zc,Geyer:2014fka} and we review briefly in \cref{sec:scattEqns}. However the functions on the moduli space which determine the states and interactions look very different in the CHY and twistor representations. The two main contributions of this paper are that we
 \begin{itemize}
 \item advance the understanding of the translation between the CHY and (ambi-)twistor representations by demonstrating and proving the splitting of general holomorphic correlators on the Riemann sphere of spinors valued in the tangent bundle of Minkowski space $T \mathbb{M}$, (which appear e.g. in the RNS string or ambi-twistor string,) into correlators of spinors valued in the spin bundles $\mathbb{S}^+$ and $ \mathbb{S}^-$. All kinematical CHY integrands (for gravity, EYM etc.) are special cases or limits of this general type of correlator.
 \item find a formula for all tree level scattering amplitudes, in all trace sectors, in  $4d$ \Neqfour supersymmetric Einstein-Yang-Mills theory.
 \end{itemize}
 This paper extends the known formulas for purely bosonic states \citep{Cachazo:2014xea}, previous work on $4d$ formulas for EYM of \citep{Adamo:2015gia}, which gave a formula for all single trace amplitudes, and the translation of CHY Pfaffians \cite{Zhang:2016rzb}. Both results are each interesting in their own right. Chiral splitting can be stated as the splitting of fermion correlators of the form
 \begin{equation}\label{eqn:fermionCorrelatorExample}
  \left \langle \prod_i (\lambda_i \lambdat_i ) \cdot \psi(z_i) \right\rangle ~, \qquad S= \int _{\prco^1} \eta_{\mu\nu} \  \psi^\mu(z) \partialb \psi^\nu(z)
  \end{equation} 
  where $(\lambda_i \lambdat_i)_\mu$ is a $ 4d$ null vector in spinor helicity notation and $\psi^\mu(z)$ is a left-moving fermionic spinor on the Riemann sphere, into two factors, each only involving the left handed $\lambda_i$ and right handed $\lambdat_i$ respectively. While chiral splitting is a general property of the correlators \eqref{eqn:fermionCorrelatorExample}, rooted in the fact that the tangent bundle of Minkowski space splits as
\begin{equation}
  T \mathbb{M} \simeq \mathbb{S}^+ \otimes  \mathbb{S}^-
  \end{equation}  
  into the left and right handed spin bundles, the present paper is interested in this because of their role in the CHY formulae. Indeed, all kinematic Pfaffians appearing in the CHY arise as correlators of this type. This means that the chiral splitting of the worldsheet correlators \cref{eqn:fermionCorrelatorExample} on the sphere lifts, via the scattering equations, to a chiral splitting of $4d$ quantum field theory amplitudes. In other words, all QFT amplitudes that can be described by a CHY formula will exhibit this chiral splitting in $4d$.

 \medskip
 Following the early work \cite{Bern:1999bx,Selivanov:1997ts,Selivanov:1997aq} there has been renewed interest recently in the study of Einstein-Yang-Mills amplitudes and their relation to pure Yang-Mills from the perspective of the double copy construction \cite{Chiodaroli:2017ngp,He:2016mzd}, string theory \cite{Schlotterer:2016cxa,Stieberger:2016lng} and the CHY formulae \cite{Teng:2017tbo,Fu:2017uzt,delaCruz:2016gnm,Nandan:2016pya}. We believe that the new formulas for \Neqfour EYM scattering amplitudes (\cref{eqn:fullamplitudeAmbiTwistorspace,eqn:fullamplitudeTwistorspace}) can provide a new tool to study these relations, particularly in light of the $4d$ KLT and BCJ relations \cite{Cachazo:2016sdc}.

 \bigskip
 We begin by very briefly reviewing the $4d$ scattering equations in \cref{sec:scattEqns}. In \cref{sec:Splitting} we discuss the technicalities and give examples of the splitting of CHY type Pfaffians, which we prove in \cref{sec:proof}, and in \cref{sec:EYMAmplitude} we present the $4d$ scattering amplitude for \Neqfour EYM on (ambit)twistor space. Both \cref{sec:Splitting,sec:EYMAmplitude} are largely self-contained, so the reader may skip directly to her/his point of interest.

 \subsection{Representations of Scattering Equations}\label{sec:scattEqns}
 It is well known that in four dimensions the scattering equations split into R-charge sectors, also known as $N^{k-2}$MHV sectors. These sectors are labelled by an integer $k$, or $d \equiv k-1$, or $\dt \equiv n - k-1$, where $n$ is the total number of particles. There are many representations of these refined scattering equations, and we will now briefly recall the three we use for this paper. The idea is to write the particles' momenta as matrices in spinor-helicity notation, and then solve the scattering equations\footnote{Since $P$ is a meromorphic $(1,0)$-form, $\det P =0$ contains $n-3$ independent equations}
 \begin{equation}
  \det P = 0  ~,
 \end{equation}
 where
 \begin{equation}
 P^{\alpha \alphad} (z) := \sum_{i\in \mathrm p} \lambda_i^\alpha \lambdat_i^\alphad ~ \frac{\diffd z}{z-z_i} ~,
 \end{equation}
 by factorizing \cite{Witten:2004cp} it as
 \begin{equation}
 P^{\alpha \alphad}(z) = \lambda ^\alpha (z ) \, \lambdat^\alphad (z) ~,
 \end{equation}
 globally on the sphere. The factorization involves a choice of how to distribute the zeros of $P$ among the two factors, and this choice labels the different refinement sectors. It also requires a choice of how to distribute the poles of $P$ among the two factors and this choice labels the various equivalent representations of the scattering equations.

  The first representation of the refined scattering equation is given by the splitting
 \begin{equation}
 P(z)  = \lambda_T(z) \, \lambdat_T(z)  
 \end{equation}
 with
\begin{equation}
 \lambda_T ^\alpha    \in H^0 \left( \mathcal{O}(d) \right) ~, \qquad \lambdat_T  ^\alphad  \in H^0 \left( \mathcal{O}(-d) \otimes K  \bigg[\sum _{i \in \mathrm p} z_i \bigg] \right) ~,
 \end{equation} 
 where the subscript stands for \emph{twistor}. The notation here means that for $\alpha = 0,1$, $\lambda_T^\alpha$ is a holomorphic polynomial of degree $d$ while $\lambdat_T^\alphad$ is a meromorphic $(1,0)$-form of homogeneity $-d$ with simple poles at all marked points. In these variables the scattering equations read
  \begin{equation}
  \res_{z_i} \lambdat_T = t_i \, \lambdat_i ~, \qquad t_i \, \lambda_T(z_i) = \lambda_i \qquad \forall i \in \mathrm  p  = \{1, \cdots ,n \} \,.
  \end{equation}
 They fix the sections $\lambda_T, \lambdat_T$, the scaling parameters\footnote{Both $\lambda_T(z)$ and $\lambda_i$ are only defined up to rescaling by a non-zero complex number. Hence the scattering equations can only require them to be proportional, and the scaling parameters $t_i$ are introduced to account for the rescaling covariance.} $t_i$ and locations $z_i$ (up to M\"obius invariance), and also enforce momentum conservation. The distinct refinement sectors are labelled by the integer $k = d+1$, and the original scattering equations $P^2=0$ are equivalent to the union of the refined scattering equations for $k = 1 , \cdots , n-1$.

  The second representation is the parity conjugate of the previous one and is given by the splitting
 \begin{equation}
 P(z)  = \lambda_{\tilde{T}}(z) \, \lambdat_{\tilde{T}}(z)  
 \end{equation}
 with
\begin{equation}
 \lambda_{\tilde{T}} ^\alpha    \in H^0 \left(   \mathcal{O}(-\dt) \otimes K  \bigg[\sum _{i \in\mathrm  p} z_i \bigg]     \right) ~, \qquad \lambdat_{\tilde{T}}^\alphad  \in H^0 \left( \mathcal{O}(\dt) \right) ~,
 \end{equation} 
 where the subscript stands for \emph{dual twistor}. Here $\lambda_{\tilde T}$ is a meromorphic $(1,0)$-form of homogeneity $-\dt$ with simple poles at all marked points, while $\lambdat_{\tilde T}$ is a holomorphic polynomial of degree $\dt$. In these variables the scattering equations read
  \begin{equation}
  \res_{z_i} \lambda_{\tilde{T}} = \tilde{t}_i \, \lambda_i ~, \qquad \tti_i \, \lambdat_{\tilde{T}}(z_i) = \lambdat_i \qquad \forall i \in \mathrm  p  
  \end{equation}
  and they again fix the sections $\lambda_{\tilde{T}}, \lambdat_{\tilde{T}}$, the scaling parameters $\tti_i$ and locations $z_i$ (up to M\"obius invariance) and enforce momentum conservation.

  The third representation is useful if there is a natural splitting of the set of external particles $\mathrm p$ into two subsets $\mathrm{p}^+ \cup \mathrm{p}^- = \mathrm p $. Then we can require that
 \begin{equation}
 P(z)  = \lambda_A(z) \, \lambdat_A(z)  
 \end{equation}
 with
\begin{equation}
 \lambda_A  ^\alpha  \in H^0 \left( K^{1/2} \bigg[\sum _{i \in \mathrm{p}^-} z_i \bigg] \right) ~, \qquad \lambdat_A^\alphad  \in H^0 \left( K^{1/2} \bigg[\sum _{i \in \mathrm{p}^+} z_i \bigg] \right) ~,
 \end{equation} 
  where the subscript stands for \emph{ambi--twistor}. Here $\lambda_A ,\lambdat_A$ are both meromorphic $(\sfrac{1}{2},0)$-forms of homogeneity $0$ and have simple poles at the marked points in $\mathrm{p}^- , \mathrm{p}^+$ respectively.  In these new variables the scattering equations read
  \begin{equation}
  \begin{aligned}
  \res_{z_i} \lambda_A = \tilde{u}_i \, \lambda_i ~, \qquad \tilde{u}_i \, \lambdat_A(z_i) = \lambdat_i \qquad \forall i \in \mathrm{p}^-  \\
  \res_{z_i} \lambdat_A = u_i \, \lambdat_i ~, \qquad u_i \, \lambda_A(z_i) = \lambda_i \qquad \forall i \in \mathrm{p}^+
  \end{aligned}
  \end{equation}
  and they fix the sections $\lambda_A, \lambdat_A$, the scaling parameters $u_i , \tilde{u}_i$ and locations $z_i$ (up to M\"obius invariance) and enforce momentum conservation.

 We can easily switch between these three representation via the relations
 \begin{equation}
 \begin{aligned}
 \lambda_T (z) ~\propto~ \frac{\prod_{i\in \mathrm{p}^-} (z-z_i)}{\sqrt{\diffd z}} ~ \lambda_A(z)  ~\propto~ \frac{\prod_{i\in \mathrm  p} (z-z_i)}{\diffd z} ~ \lambda_{\tilde T}(z) \\
 \lambdat_T (z) ~\propto~ \frac{\sqrt{\diffd z}}{\prod_{i\in \mathrm{p}^-} (z-z_i)} ~ \lambdat_A(z)  ~\propto~ \frac{\diffd z}{\prod_{i\in \mathrm p} (z-z_i)} ~ \lambdat_{\tilde T}(z)
 \end{aligned}
 \end{equation}
 for the sections, where the factor of proportionality is independent of $z$, and also for the scaling parameters
 \begin{equation}
 \frac{t_j}{t_i} \, \prod_{k \in \mathrm{p}^- \backslash  \{ i\} } \frac{z_j - z_k}{z_i - z_k}   = \tilde{u}_i \, u_j  ~ \frac{\sqrt{\diffd z_i \, \diffd z_j}}{z_i-z_j}= \frac{\tti_i}{\tti_j} \, \prod_{k \in \mathrm{p}^+ \backslash  \{ j\} } \frac{z_i - z_k}{z_j - z_k} 
 \end{equation}
 for any choice of $    ~ i \in \mathrm{p}^- , j \in \mathrm{p}^+$. The locations $z_i$ are identical among the three representations.
 
 Notice that among the three representations the number of zeros in $\lambda_A, \lambda_T , \lambda_{\tilde T}$ and $\lambdat_A, \lambdat_T , \lambdat_{\tilde T}$ is always $d$ and $\dt$ respectively, and only the poles are redistributed. Of course one may define many more representations of the same equations by choosing different ways of distributing the poles among the two factors, but for the present paper we will only need these three.

 \section{Chiral Splitting of Fermion Correlators \& CHY Pfaffians}\label{sec:Splitting}
 The key step in the translation of the CHY integrands into spinor--helicity language is the factorization of the kinematic Pfaffians into Hodges matrices \cite{Hodges:2012ym,Cachazo:2012kg}. Take $2n$ points on the sphere $z_i$ and to each point associate one un-dotted (left-handed) spinor $\lambda_i$ and one dotted (right-handed) spinor $\lambdat _i$. The basic identity which we found and use throughout the rest of the paper is the factorization of the Pfaffian
 \begin{equation}\label{eqn:pfaffianFactorizationProto}
 \pf \left( \frac{\itw{\lambda_i}{\lambda_j} \ditw {\lambdat_i}{\lambdat_j}}{z_i - z_j} \right)^{i,j=1,\cdots , 2n} =  ~\frac{\, \det \left( \frac{\itw {\lambda_i}{\lambda_j}}{z_i - z_j} \right)^{i\in \mathrm{b}}_{j\in \mathrm{b}^c} }{V( \mathrm{b}) \, V( \mathrm{b} ^c)}   ~~ \frac{\,  \det \left( \frac{\ditw {\lambdat_i}{\lambdat_j}}{z_i - z_j} \right)^{i\in \tilde{ \mathrm{b}}}_{j\in \tilde{ \mathrm{b}}^c} }{V( \tilde{ \mathrm{b}}) \, V( \tilde{ \mathrm{b}} ^c)} ~~ V( \{ 1 , \cdots , 2n \} )
 \end{equation}
 where $ \mathrm{b}, \tilde{ \mathrm{b}}$ are arbitrary ordered\footnote{The expression \cref{eqn:pfaffianFactorizationProto} is easily seen to be independent of the ordering of $ \mathrm{b}, \tilde{ \mathrm{b}}$, but the Hodges determinant and the Vandermonde determinant separately are not, so we keep track of the ordering.} subsets of $\{1, \cdots , 2n \}$ of size $n$ and $\mathrm{b}^c , \tilde{ \mathrm{b}} ^c$ are their complements. We use the notation that $\det(M_{ij})^{i\in \mathrm a}_{j\in \mathrm b}$ denotes the determinant of the matrix $M$, with rows indexed by the set $\mathrm a$ and columns by the set $\mathrm b$. Since the Pfaffian is only defined for antisymmetric matrices, it's rows and columns are necessarily indexed by the same set. We also use the Vandermonde determinant, defined as usual
 \begin{equation}
 V( \mathrm{b} ) = \prod_{i<j \in \mathrm{b}} (z_i - z_j) 
 \end{equation}
for an ordered set of points on the sphere. It is worth emphasizing that this factorization does not require the scattering equations to hold (and that the spinors $\lambda _i , \lambdat_i$ need not have any interpretation in terms of null momenta or polarization vectors, though of course that is how we will employ this formula below). The kinematic Pfaffians for gravity, EYM, etc. may all be realized as appropriate limits or special cases of this Pfaffian.

To prove \cref{eqn:pfaffianFactorizationProto} we simply compute the residues as any $z_i - z_j \to 0$ on both sides and invoke induction. We outline the idea of the proof here and point the interested reader to \cref{sec:proof} for details: At first glance it seems as though the right hand side depends on the splitting of the $2n$ points into the two halves $\mathrm{b} , \mathrm{b} ^c$ and $\tilde{ \mathrm{b}} , \tilde{ \mathrm{b}} ^c$ respectively, which would be at odds with the manifest $S_{2n}$ antisymmetry of the Pfaffian on the left. This tension is resolved by the surprising fact that the combination
 \begin{equation}\label{eqn:buildingblock}
 \frac{\, \det \left( \frac{\itw ij}{z_i - z_j} \right)^{i\in \mathrm{b}}_{j\in \mathrm{b}^c}  }{V( \mathrm{b}) ~ V( \mathrm{b} ^c)}  
 \end{equation}
 is totally $S_{2n}$ permutation symmetric, despite making only the permutation invariance under a $S_n \times S_n \times \mathbb{Z}_2$ subgroup manifest. To exhibit full permutation invariance we may go to an alternative representation
 \begin{equation}\label{eqn:buildingBlockTwistinvariance}
 \frac{\, \det \left( \frac{\itw ij}{z_i - z_j} \right)^{i\in  \mathrm{b} }_{j\in \mathrm{b}^c } }{V(   \mathrm{b} ) ~ V(   \mathrm{b} ^c )}~~   = (-1)^{\frac{n \,(n-1)}{2}} \sum_{\substack{\mathrm{p}      \subset \{ 1 , \cdots , 2n \}     \\ |\mathrm{p}| =n  }} \frac{   \prod_{i \in \mathrm{p} }  (\lambda_i) ^0   ~   \prod_{j \in \mathrm{p}^c }  (\lambda_j) ^1   }{   \prod_{\substack{i \in \mathrm{p} \\ j \in \mathrm{p}^c }} (z_i - z_j)  }
 \end{equation}
 where the sum runs over all unordered subsets $\mathrm{p} \subset \{ 1 , \cdots , 2n \}$ of size $n$. The right hand side is now manifestly $S_{2n}$ permutation invariant (though it has lost its manifest $SL(2)$ Lorentz invariance). This last equality is interesting in its own right, but will not be used in the present paper other than to prove the $S_{2n} $ symmetry of \cref{eqn:buildingblock}. Below we will use the $S_{2n} $ symmetry of \cref{eqn:buildingblock} repeatedly in order to streamline the calculations.

 \medskip
 We now demonstrate the chiral splitting formula \cref{eqn:pfaffianFactorizationProto} by translating various CHY formulae into 4d spinor helicity variables.

 \subsection{Refinement of the Scalar Mode Pfaffian}
 As a warm-up we demonstrate how to use \cref{eqn:pfaffianFactorizationProto} to factorize the scalar mode CHY Pfaffian
 \begin{equation}
 \pf (A) = \pf \left( \frac{p_i \cdot p_i }{ z_i  -z_j } \right)^{i,j=1, \cdots , n} = \pf \left( \frac{\itw{\lambda_i}{\lambda_j} \ditw {\lambdat_i}{\lambdat_j}}{z_i - z_j} \right)^{i,j=1,\cdots ,  n}
 \end{equation}
 with $n$ even. Now we have to choose two ways of splitting of the $n $ labels into two halves, one for the rows and one for the columns respectively. Choosing, for example, the splitting $1 ,\cdots ,n/2$ and $n/2+1 , \cdots , n $ for both for both angle and square brackets, by \cref{eqn:pfaffianFactorizationProto}  we find
 \begin{equation}
 \begin{aligned}
 \pf (A) = ~&   \det \left( \frac{\itw {\lambda_i}{\lambda_j}}{z_i - z_j} \right)^{i = 1 , \cdots , n/2}_{j = n/2+ 1 , \cdots , n}   ~\cdot ~  \det \left( \frac{\ditw {\lambdat_i}{\lambdat_j}}{z_i - z_j} \right)^{i = 1 , \cdots , n/2}_{j = n/2+ 1 , \cdots , n}  \\
  & ~~\cdot \frac{ \prod_{i = 1 }^{n/2}  \prod_{j = n/2+1 }^n (z_i  -z_j) }{    \prod_{i,j = 1 }^{n/2}   (z_i  -z_j)   ~    \prod_{i,j = n/2+1 }^n (z_i  -z_j)  }
 \end{aligned}
 \end{equation}
 Of course this is just one way to chose the distribution of the $n $ row/column labels of the Pfaffian onto the two determinants, and they're all equivalent (after taking into account the appropriate Vandermonde ratios).
 
 Since $\pf(A)$ has corank $2$ when evaluated on solutions to the scattering equations \cite{Cachazo:2014xea}, the above Pfaffian actually vanishes and we instead consider the reduced Pfaffian, defined to be the Pfaffian of any $n-2 \times n-2$ minor of $A$, together with a Jacobian factor which preserves the $S_n$ permutation invariance of the construction. We can adapt the above easily
 \begin{equation}
 \begin{aligned}
 \pf ^\prime (A)  = ~& \frac{1}{z_1 - z_n } ~ \pf \left( \frac{\itw{\lambda_i}{\lambda_j} \ditw {\lambdat_i}{\lambdat_j}}{z_i - z_j} \right)^{i,j=2,\cdots ,  n-1} \\
 = ~&   \det \left( \frac{\itw {\lambda_i}{\lambda_j}}{z_i - z_j} \right)^{i = 2 , \cdots , n/2}_{j = n/2+ 1 , \cdots , n-1}   ~\cdot ~  \det \left( \frac{\ditw {\lambdat_i}{\lambdat_j}}{z_i - z_j} \right)^{i = 2 , \cdots , n/2}_{j = n/2+ 1 , \cdots , n-1}  \\
  & ~~ \cdot \frac{1}{z_1 - z_n }  ~\frac{ \prod_{i = 2 }^{n/2}  \prod_{j = n/2+1 }^{n-1} (z_i  -z_j) }{    \prod_{i,j = 2 }^{n/2}   (z_i  -z_j)   ~    \prod_{i,j = n/2+1 }^{n-1} (z_i  -z_j)  } ~.
 \end{aligned}
 \end{equation}
 Again, for concreteness, we display just one of many equivalent ways of splitting the reduced Pfaffian into two determinants. Notice that the splitting does not require the scattering equations to hold, but on the support of the scattering equations the above expression becomes $S_n$ symmetric.

 \subsection{Refinement of the Vector--Mode Pfaffian}
 
 As next example we shall translate the probably best known CHY integrand, the kinematic Pfaffian for massless vector modes \cite{Cachazo:2013hca}
 \begin{equation}\label{eqn:vectorPfaffian}
 \pf \left( \begin{array}{cc}
A & -C^T \\  C & B 
\end{array}  \right)
 \end{equation}
 with the entries of each block-matrix given as
 \begin{equation}
 A_{ij} = p_i \cdot p_j \,S( z_i , z_j ) ~,\qquad   B_{ij} = \varepsilon_i \cdot \varepsilon_j \,S( z_i , z_j )  ~,\qquad   C_{ij} = \varepsilon_i \cdot p_j \,S( z_i , z_j ) 
 \end{equation}
 and 
 \begin{equation}
 A_{ii} =0 ~, \qquad  B_{ii} = 0  ~,  \qquad  C_{ii} = \varepsilon_i \cdot  P(z_i) ~,
 \end{equation}
 where
 \begin{equation}
 S(z , w) := \frac{\sqrt{  \diffd z \, \diffd w} }{z-w}
 \end{equation}
 is the free fermion propagator (Szeg\'o kernel) on the Riemann sphere\footnote{The factors of $\sqrt{\diffd z}$ can be removed using the multilinearity of the Pfaffian, but we keep them in place to highlight its CFT origin.}. It is convenient to use the following parametrization for polarization vectors 
 \begin{equation}
 \varepsilon^- _i = \frac{ \nket{ \lambda_i } \pbra {\xi_i} }{\ditw{\xi_i }{\lambdat_i}}  ~, \qquad   \varepsilon^+_i = \frac{\nket {\xi_i}  \pbra{ \lambdat_i }  }{\itw{\xi_i }{\lambda_i}}  ~,
 \end{equation}
 for states of negative and positive helicity respectively. For the following discussion we fix the degree of the refined scattering equations to be $d = |\mathrm{p}^-| -1$, for which we give a justification below.

 In order to use the factorization formula \cref{eqn:pfaffianFactorizationProto} on this Pfaffian we first have to cast it into the form
 \begin{equation}
 \pf \big( q_i \cdot q_j \, S(z_i , z_j) \big) ^{i,j = 1 , \cdots ,2n} ~.
 \end{equation}
 We would like to identify $q_i = p_i$ and $q_{i+n} = \varepsilon_i$ as well as $z_i = z_{i+n}$ for $i = 1 , \cdots , n$, but the diagonal terms in the block $C$ present an obstruction to doing so. We can resolve this obstruction by employing a point splitting procedure and using the scattering equations. The idea is to introduce $n$ new marked points on the sphere, one $w_i$ associated to each $z_i$, and write the momenta as
 \begin{equation}
 p_i = \lim _{w_i \to z_i } p_i(w_i)    \equiv \lim _{w_i \to z_i } \left\{  t_i \, \tti_i ~ \nket{ \lambda_T(w_i) } \pbra{ \lambdat_{\tilde T} (w_i) } \right\}
 \end{equation}
 and also write
  \begin{equation}
 \varepsilon^-_i   =  t_i ~  \frac{ \nket{ \lambda_T (z_i) } \pbra {\xi_i} }{\ditw{\xi_i }{\lambdat_i}}   \qquad \text{and} \qquad \varepsilon^+_i   =  \tti_i   ~  \frac{ \nket{ \xi_i } \pbra{ \lambdat_{\tilde T}(z_i) }}{\itw{\xi_i }{\lambda_i}}   ~.
 \end{equation}
  Here we use the functions $\lambda_T $ and $\lambdat_{\tilde T} $ from the twistor and dual twistor representation of the refined scattering equations, respectively. We shall work in this enlarged description\footnote{In the language of the Ambitwistor String model of \cite{Mason:2013sva} this means that we write the descended vertex operators as a product
 \begin{equation}
 \varepsilon_i \cdot P(z_i) + : \varepsilon_i \cdot \psi (z_i) ~ p_i \cdot \psi (z_i) : ~=~ \lim_{w_i \to z_i} p_i(w_i) \cdot\psi(w_i)  ~ \varepsilon_i \cdot \psi(z_i) ~.
 \end{equation}
 The correlator of these point--split vertex operators gives rise to the Pfaffian in \cref{eqn:pointsplitPfaffian}.} to facilitate the factorization of the Pfaffian, and take the limit $w_i \to z_i$ only at the very end, where we recover the original momentum vectors. The upshot is that the diagonal terms of $C$ may now be written as
 \begin{equation}
 \varepsilon^\pm_i \cdot P(z_i)  =   \lim_{w_i \to z_i} \bigg( \varepsilon_i ^\pm \cdot p_i(w_i)  ~ S(z_i , w_i)  \bigg)
 \end{equation}
 and hence we have succeeded in bringing the Pfaffian into the desired form
 \begin{equation}\label{eqn:pointsplitPfaffian}
  \pf \left( \begin{array}{cc}
A & -C^T \\  C & B 
\end{array}  \right) = \lim_{w_i \to z_i}   \pf \left( \begin{array}{cc}
p_i(w_i) \cdot p_j(w_j) \, S(w_i ,w_j) ~&~ p_i(w_i) \cdot \varepsilon_j  \, S(w_i,z_j) \\  \varepsilon_i  \cdot p_j(w_j) \, S(z_i ,w_j)  \,  & \varepsilon_i  \cdot \varepsilon_j  \, S(z_i,z_j) 
\end{array}  \right)  ~.
 \end{equation}
 Clearly, the only non-trivial part of this statement is that the diagonal terms in $C $ indeed have the correct limit, which we demonstrate below.

\bigskip
 Having brought the Pfaffian into the canonical form \cref{eqn:pointsplitPfaffian} we may now use \cref{eqn:pfaffianFactorizationProto} to factorize it and find
 \begin{equation}
 \begin{aligned}
    &  \pf \left( \begin{array}{cc}
p_i(w_i) \cdot p_j(w_j) \, S(w_i ,w_j) ~&~ p_i(w_i) \cdot \varepsilon_j  \, S(w_i,z_j) \\  \varepsilon_i  \cdot p_j(w_j) \, S(z_i ,w_j)  \,  & \varepsilon_i  \cdot \varepsilon_j  \, S(z_i,z_j) 
\end{array}  \right)   \\
  = & ~ \det \left( \begin{array}{cc}
 \itw{\lambda(z_i)}{\lambda(w_j) } \, S(z_i,w_j) \\  \itw{ \xi_k} {\lambda(w_j) }\, S(z_k,w_j) 
\end{array}  \right)    ^{i \in  \mathrm{p}^- , \, k \in \mathrm{p}^+}_{j \in \mathrm{p}^- \cup \mathrm{p}^+}    ~~~  \cdot \prod_{i\in \mathrm{p}^+} \frac{1}{S(w_i,z_i)} \frac{\tti_i^2}{\itw {\lambda_i}{\xi_i}}  \\
&{} \cdot \det \left( \begin{array}{cc}
    \ditw{\xi_i}{  \lambdat(w_j) }\, S(z_i,w_j) \\ \ditw{\lambdat(z_k)}{\lambdat(w_j) } \, S(z_k,w_j) 
\end{array}  \right)    ^{i \in  \mathrm{p}^- , \, k \in \mathrm{p}^+}_{j \in \mathrm{p}^- \cup \mathrm{p}^+}    ~~~   \cdot \prod_{i\in \mathrm{p}^-} \frac{1}{S(w_i,z_i)}    \frac{t_i ^2}{\ditw {\lambdat_i}{\xi_i}}    \\
& {} \cdot \frac{\prod_{i\neq j =1}^n (z_i - w_j)}{\prod_{i< j =1}^n (z_i - z_j)(w_i - w_j)} 
 \end{aligned}
 \end{equation}
 where we have chosen the splitting such that the rows are labelled by the punctures $z_i$ while the columns are labelled by the $w_i$. The sub--blocks of both determinants are of size $|\mathrm{p}^-| \times n$ and $|\mathrm{p}^+| \times n$ respectively.
 
 The last step is to take the limit $w_i \to z_i$. Notice that each line is finite in the limit, as the potential singularities on the diagonal of the determinants are cancelled by the inverse Szego kernels multiplying them. Explicitly, we see for example that the first line becomes
 \begin{equation}\label{eqn:pointsplitLimitHodgesMatrix}
 \lim _{w_i \to z_i}  \det \left( \begin{array}{cc}
 \itw{\lambda(z_i)}{\lambda(w_j) } \, S(z_i,w_j) \\  \itw{ \xi_k} {\lambda(w_j) }\, S(z_k,w_j) 
\end{array}  \right)    ^{i \in  \mathrm{p}^- , \, k \in \mathrm{p}^+}_{j \in \mathrm{p}^- \cup \mathrm{p}^+}    ~~~  \cdot \prod_{i\in \mathrm{p}^+} \frac{1}{S(w_i,z_i)} \frac{1}{\itw {\lambda_i}{\xi_i}}  =   \det (\Phi)^{i\in \mathrm{p}^-}_{j\in \mathrm{p}^-}
 \end{equation}
 where we recover the dual Hodges matrix with entries
 \begin{equation}
 \Phi_{ij} = \itw{\lambda(z _i) }{\lambda(z_j)} \, S(z_i ,z_j) ~,\qquad \text{and} \qquad \Phi_{ii} =  \itw{\lambda (z_i) }{\diffd \lambda(z_i)}  ~.
 \end{equation}
 To take the limit $w_i \to z_i$ in \eqref{eqn:pointsplitLimitHodgesMatrix} we used the Leibniz formula for the determinant and then noticed that the terms surviving in the limit reassemble into $\det ( \Phi )$. The second line similarly gives rise to the determinant of the Hodges matrix with entries
 \begin{equation}
 \Phit_{ij} = \ditw{\lambdat(z _i) }{\lambdat(z_j) } \, S(z_i ,z_j) ~,\qquad \text{and} \qquad \Phit_{ii} = \ditw{\lambdat(z _i) }{ \diffd\lambdat(z_j) }    ~.
 \end{equation}
 Recall that the polynomials $\lambda(z), \lambdat(z)$ belong to the twistor/dual twistor representation of the refined scattering equations, respectively, so these definitions of the Hodges matrices agree with the usual ones on the support of the scattering equations (of the correct degree).
 
 \bigskip
To summarize, we have shown that the vector--mode CHY Pfaffian factorizes into a product of a Hodges determinant times a dual Hodges determinant
\begin{equation}\label{eqn:vectorPfaffianFactorizationFinal}
 \pf \left( \begin{array}{cc}
A & -C^T \\  C & B 
\end{array}  \right) ~=~ \det(\Phi) ^{i \in \mathrm{p}^-}_{j \in \mathrm{p}^-} ~ \cdot ~ \det(\Phit) ^{i \in \mathrm{p}^+}_{j \in \mathrm{p}^+}  \cdot \prod_{i\in \mathrm{p}^-} t_i^2  \cdot \prod_{j\in \mathrm{p}^+} \tti_j^2
\end{equation}
 on the support of the scattering equations of degree $d = |\mathrm{p}^-| -1$.

 \subsubsection{Degree, Kernel and $C_{ii}$ Diagonal Elements}
 There are several loose ends to tie up in the above discussion. Firstly, we have used the refined scattering equations of degree $d = |\mathrm{p}^-| -1$ without justification for fixing the degree of the scattering equations in terms of the number of particles with negative helicity. Indeed, the CHY scattering equations are equivalent to the union of the refined scattering equations of all possible degrees, so a priori there is no reason to restrict our attention to the refined scattering equations in the sector $k = |\mathrm{p}^-| $ only. It is however known \cite{Zhang:2016rzb} that the Pfaffian \cref{eqn:vectorPfaffian} actually vanishes when evaluated on solutions to the refined scattering equations of the wrong degree. One way to show this is to perform the same steps as above\footnote{In the twistor and dual twistor representation the diagonal terms in the C matrix block remain of the same form even when $d \neq |\mathrm{p}^- | - 1$. See below for further comments.} and then discover that one of the Hodges matrices has a larger than expected kernel. Though this is straightforward, we want to take an alternative route here.
 
 We can actually construct the kernel of the CHY--matrix evaluated on the wrong degree explicitly. In fact, if $0 < \Delta := |\mathrm{p}^- | -1 -d $, then define
 \begin{equation}
 v_ i = \gamma(z_i ) \, t_i^{-1} \,  \frac{[\xi _i  | \res_{z_i} \tilde{\zeta}  ]}{ [ \xi_i \, \lambdat_i ]} ~, \qquad w_ i = -\gamma(z_i ) \, t_i^{-1} \,  [\lambdat_i | \res_{z_i} \tilde{\zeta}  ]  \qquad \text{for } i = 1 , \cdots ,n
 \end{equation}
 where $\gamma \in H^0( T^{1/2})$ is any holomorphic section of $T^{1/2}$ and
 \begin{equation}
 \tilde{\zeta}^{\dot \alpha} \in H^0 \left( \mathcal{O}(-d) \otimes K \bigg[\sum_{i=1}^n z_i \bigg] \right)
 \end{equation}
 with the requirement that
 \begin{equation}
 \res_{z_i}  \zeta  = t_i \, \lambdat_i \qquad \forall i \in \mathrm{p}^+ ~.
 \end{equation}
 Here $\xi_i $ are the auxiliary spinors that enter the definition of the polarization vectors when $i \in \mathrm{p}^-$, and arbitrary spinors when $i \in \mathrm{p}^+$. (Note that this requirement implies some simplifications of the kernel, e.g. $w_i = 0$ for $i \in \mathrm{p}^+$. Also we find that under a gauge transformation $ \varepsilon_i \to \varepsilon_i + p_i $ the kernel transforms as $v_i \to v_i - w_i $, which is necessary for the following equation to be gauge covariant.) With these definitions a straightforward, though somewhat tedious, calculation shows that the scattering equations imply
 \begin{equation}
 \left( \begin{array}{cc}
A & -C^T \\  C & B 
\end{array}  \right) \left( \begin{array}{c}
v \\ w
\end{array}  \right) = 0
 \end{equation}
 Counting the free parameters in $\zeta$ and $\gamma$ we find that the kernel is of dimension $2 \, \Delta +2$.
 
 If $\Delta <0 $ we have to take the parity conjugate of the above construction and find that the kernel is of dimension $2 \, |\Delta| +2$.

 \bigskip
 The second loose end to tie up is that even when evaluated on solutions of the correct degree, the Pfaffian and the Hodges matrices have a non--empty kernel. Hence, the above discussion needs to be adapted to the reduced Pfaffian
 \begin{equation}
  \pf ^\prime \left( \begin{array}{cc}
A & -C^T \\  C & B 
\end{array}  \right)   = S(z_1 , z_2) ~  \pf  \left( \begin{array}{cc}
A & -C^T \\  C & B 
\end{array}  \right)   ^{\check 1, \check 2}
 \end{equation}
 where the superscript $\check 1, \check 2 $ is the instruction to remove the first two rows and columns from the matrix before taking its Pfaffian. The scattering equations ensure that the reduced Pfaffian is still fully permutation symmetric, albeit not manifestly so. We may assume without loss of generality that $1 \in \mathrm{p}^-$ and $2 \in \mathrm{p}^+$. Then, retracing the steps from above with one fewer row/column in each matrix, we find 
 \begin{equation}
 \begin{aligned}
      \pf ^\prime  \left( \begin{array}{cc}
A & -C^T \\  C & B 
\end{array}  \right) & \\
  = & ~ \det \left( \begin{array}{cc}
 \itw{\lambda(z_i)}{\lambda(z_j) } \, S(z_i,z_j) \\  \itw{ \xi_2} {\lambda(z_j) }\, S(z_2,z_j) 
\end{array}  \right)    ^{i \in  \mathrm{p}^- \backslash \{1\} }_{j \in \mathrm{p}^- }    ~~~ \cdot \frac{ 1}{\itw {\lambda_2}{\xi_2} \, \tti_2}  \cdot \prod_{i\in \mathrm{p}^+}   \tti_i^2  \\
&{} \cdot \det \left( \begin{array}{cc}
    \ditw{\xi_1}{  \lambdat(z_j) }\, S(z_1,z_j) \\ \ditw{\lambdat(z_k)}{\lambdat(z_j) } \, S(z_k,z_j) 
\end{array}  \right)    ^{  k \in \mathrm{p}^+ \backslash \{2\} }_{j  \in \mathrm{p}^+}    ~~~  \cdot \frac{ 1}{\ditw {\lambdat_1}{\xi_1} \, t_1}  \cdot \prod_{i\in \mathrm{p}^-}   t_i^2 
 \end{aligned}
 \end{equation}
where the two matrices each consist of two sub-blocks, with dimensions $1 \times d+1$ and $d \times d+1$ and $1 \times \dt+1$ and $\dt \times \dt +1$ respectively. To bring this into the desired form of two reduced Hodges determinants we use that one can add linear combinations of the columns in a matrix onto each each other without changing the value of the determinant. Thus we can show that for instance
\begin{equation}
\begin{aligned}
   \det \left( \begin{array}{cc}
 \itw{\lambda(z_i)}{\lambda(z_j) } \, S(z_i,z_j) \\  \itw{ \xi_2} {\lambda(z_j) }\, S(z_2,z_j) 
\end{array}  \right)  ^{i \in  \mathrm{p}^- \backslash \{1\} }_{j \in \mathrm{p}^- }   = \,& \det \left( \begin{array}{cc}
 \itw{\lambda(z_i)}{\lambda(z_j) } \, S(z_i,z_j) \\  \itw{ \xi_2} {\lambda(z_2) }\, \delta_{1,j} 
\end{array}  \right)  ^{i \in  \mathrm{p}^- \backslash \{1\} }_{j \in \mathrm{p}^-   }  \\
 {}&{} \cdot   S(z_1,z_2) \prod_{k \in \mathrm{p}^- \backslash \{1\}}\frac{z_1 - z_k}{z_2-z_k}    \\
 {}= \,& \det \left( \Phi \right)  ^{i \in  \mathrm{p}^- \backslash \{1\} }_{j \in \mathrm{p}^- \backslash \{1\} }  ~~ \cdot ~~ \itw{ \xi_2 }{ \lambda (z_2)}  \\
  &{} \cdot   S(z_1,z_2) \prod_{k \in \mathrm{p}^- \backslash \{1\}}\frac{z_1 - z_k}{z_2-z_k}  
\end{aligned}
\end{equation}
where we added the columns for $j \in \mathrm{p}^- \backslash \{1\}$ with coefficients
\begin{equation}
 \frac{\sqrt{\diffd z_1}}{\sqrt { \diffd z_j }} \prod_{k \in \mathrm{p}^- \backslash \{1, j\}}\frac{z_1 - z_k}{z_j-z_k}   ~,
\end{equation}
 onto the column $j =1 $. This simplifies the determinant because the last row now has only a single non-vanishing entry. Indeed, Cauchy's theorem tells us that
 \begin{equation}
 \begin{aligned}
 \itw{ \xi _2 }{ \lambda (z_1)  } \, S(z_2,z_1) + \sum_{j \in \mathrm{p}^- \backslash \{1\} }  \itw{ \xi _2 }{ \lambda (z_j)  } \, S(z_2,z_j)   ~  \frac{\sqrt{\diffd z_1}}{\sqrt { \diffd z_j }} \prod_{k \in \mathrm{p}^- \backslash \{1, j\}}\frac{z_1 - z_k}{z_j-z_k}  \\
 =  \itw{ \xi _2 }{ \lambda (z_2)  }  \, S(z_1,z_2)  \prod_{k \in \mathrm{p}^- \backslash \{1, j\}}\frac{z_1 - z_k}{z_2-z_k} 
 \end{aligned}
 \end{equation}
 for the last row, as well as
 \begin{equation}
 \Phi_{i1} + \sum_{j \in \mathrm{p}^- \backslash \{1\} } \Phi_{ij}   ~  \frac{\sqrt{\diffd z_1}}{\sqrt { \diffd z_j }} \prod_{k \in \mathrm{p}^- \backslash \{1, j\}}\frac{z_1 - z_k}{z_j-z_k} =0
 \end{equation}
 for all rows labelled by $i \in \mathrm{p}^- \backslash \{1\}$. After the analogous argument for the Hodges matrix we find that the reduced Pfaffian factorizes as expected into two reduced Hodges determinants
 \begin{equation}
  \pf ^\prime \left( \begin{array}{cc}
A & -C^T \\  C & B 
\end{array}  \right)     =     ~ \det \left( \Phi \right)    ^{i \in  \mathrm{p}^- \backslash \{1\} }_{j \in \mathrm{p}^-\backslash \{1\} }         \cdot \det \left(\Phit \right)    ^{  k \in \mathrm{p}^+ \backslash \{2\} }_{j  \in \mathrm{p}^+ \backslash \{2\} }      \cdot \frac{1}{\tilde{u}_1^2 \, u_2^2}   \cdot \prod_{i\in \mathrm{p}^- \backslash \{1\}}   t_i^2  \cdot \prod_{i\in \mathrm{p}^+ \backslash \{2\}}   \tti_i^2 
 \end{equation}
 Recall that the scaling parameters $t_i, \tti_i , u_i ,\tilde{u}_i$ come from the various representations of the refined scattering equations in the sector $ d= |\mathrm{p}^-|-1$.

\bigskip
 Finally we would like to spell out the details pertaining to the diagonal elements of the $C$ block--matrix that were used in the point--splitting procedure above. It is known that on the support of the scattering equations the $C_{ii}$ are gauge invariant, and in fact reduce to the diagonal elements of the Hodges matrices. Indeed for a negative helicity particle, $i \in \mathrm{p}^-$, we compute
 \begin{equation}
 \epsilon_i^- \cdot P(z_i) = \lim_{z \to z_i}  \epsilon_i^- \cdot P(z) = \lim_{z \to z_i}  \itw{ \lambda_i }{ \lambda_A(z) } \, \frac{\ditw{\xi_i }{\lambdat_A(z)}}{\ditw{\xi_i }{\lambdat_i}}  = \itw{ \lambda_i }{ \lambda_A(z_i) } \, \tilde{u}_i ^{-1}
 \end{equation}
 using the scattering equations in ambi--twistor form, and likewise $\epsilon_i^+ \cdot P(z_i)   = \ditw{ \lambdat_i }{ \lambdat_A(z_i) } \, u_i ^{-1}$ for $i \in \mathrm{p}^+$. Notice that while, for instance, $\lambda_A(z)$ has a pole at $z_i$ for $i \in \mathrm{p}^-$, the combination $\itw{ \lambda_i }
 {\lambda_A(z)}$ is regular at $z \to z_i$. Analogous comments apply to $\lambdat_A(z)$ and of course $P(z)$. Using the relations between the three representations of the refined scattering equations we may write these diagonal terms equivalently as
 \begin{equation}
 \begin{aligned}
\epsilon_i^- \cdot P(z_i) &  ~=~  \itw{  \lambda_i }{ \diffd \lambda_T(z_i) } ~ t_i ~=~  \itw{  \lambda_i }{   \lambda_{\tilde T}(z_i) } ~ \tti_i^{-1}  \\
\epsilon_i^+ \cdot P(z_i) & ~=~  \ditw{  \lambdat_i }{   \lambdat_{ T}(z_i) } ~ t_i^{-1}   ~=~  \ditw{  \lambdat_i }{ \diffd \lambdat_{\tilde T}(z_i) } ~ \tti_i
\end{aligned}
 \end{equation}
 for $i \in \mathrm{p}^- \cup \mathrm{p}^+$. Note that the twistor and dual twistor representations of these terms do not rely on the fact that $d= |\mathrm{p}^-| -1$, so they can be used straightforwardly even in solution sectors with $d\neq |\mathrm{p}^-| -1$.

\subsection{Refinement of the `Squeezed' Vector--Mode Pfaffian}
 Having discussed the factorization of the vector mode Pfaffian in great detail, we may now apply the same technique to the squeezed Pfaffian appearing in the CHY formula for Einstein-Yang-Mills. Recall from \cite{Cachazo:2014xea} the half-integrand for EYM tree amplitudes in the sector with $\tau$ colour traces reads
\begin{equation}
 \pf ^\prime ( \Pi (\tr_1 , \cdots , \tr_\tau : \mathrm{h}) ) =    \sum_{\substack{i_2 < j_2 \in \tr _2 \\  \cdots \\ i_\tau < j_\tau \in \tr _\tau }}  \prod_{\alpha =2}^m (z_{i_\alpha} -z_{j_\alpha} ) ~~ \pf \left(M(  \mathrm{h}  \cup \mathrm{I} \cup \mathrm{J} : \mathrm{h})  \right)    
 \end{equation} 
 with the abbreviation for the gluon labels
\begin{equation}
\mathrm{I} \equiv \{ i_2 , \cdots , i_\tau \}   \qquad \text{and} \qquad \mathrm{J} \equiv  \{ j_2 , \cdots , j_\tau \} ~.
\end{equation}
 Note that $\pf ^\prime \Pi$ only makes explicit reference to $\tau-1$ traces. On the support of the scattering equations this reduced Pfaffian does not depend on which trace is being removed from the expression.
 
 To factorize this we may use the splitting formula \cref{eqn:pfaffianFactorizationProto} term by term in the sum. Since the formula for EYM scattering amplitudes contains also a vector mode Pfaffian for gravitons and gluons, we find using the Kernel argument from above that the amplitude localizes to solutions of degree $d = n_{gr}^- +  n_{gl}^- -1$.  For the gravitons we have to employ the point-splitting procedure as above, and the structure is identical to the pure vector Pfaffian. For the gluons we don't need to point-split and can just take the result for the scalar mode Pfaffian. Combining the two we find  
\begin{equation}
\pf \left(M(  \mathrm{h}  \cup \mathrm{I} \cup \mathrm{J} : \mathrm{h})  \right)       ~=~  \det \left( \Phi ^{i \in \mathrm{h}^- \cup I}  _{j \in \mathrm{h}^- \cup J} \right)   ~ \det \left( \Phit    ^{i \in \mathrm{h}^+ \cup I}  _{j \in \mathrm{h}^+ \cup J}\right)  ~   \frac{ V( I \cup J ) }{   V( I) ^2 V(J ) ^2 } ~,
 \end{equation} 
 with $\Phi, \Phit$ the Hodges matrices as defined above. We have chosen to split the rows/columns in a symmetric way, but again, many others are possible using \cref{eqn:pfaffianFactorizationProto}.

\section{EYM Tree Scattering Amplitudes}\label{sec:EYMAmplitude}
 Having factorized the CHY integrand for 4d EYM into two chiral halves, we can lift it to a formula for all tree--level scattering amplitudes in maximally supersymmetric Einstein-Yang-Mills.

 It is well known that $4d$ scattering amplitudes can be organized by MHV sector \cite{Parke:1986gb,Berends:1987me}, which counts the number of states of one helicity, and is independent of the number of states of the other helicity. The remarkable simplicity of the maximally helicity violating amplitudes can be traced back to the integrability properties of the underlying (anti-self-dual) field equations, and the higher $N^{k-2}$MHV amplitudes are an expansion around this integrable sector. While this perspective breaks manifest parity invariance, it retains a natural action of parity and the emergence of parity invariance is understood \cite{Witten:2004cp,Roiban:2004yf}. After incorporation of supersymmetry, the MHV sectors are generalized to $R$-charge super-selection sectors. This continues to be true in \Neqfour EYM, which is expected already from the CHY representation: Since the CHY integrand for EYM still contains one vector mode, alongside one squeezed vector mode Pfaffian, the specialization to a definite degree $d = k-1$ of the scattering equations still occurs, where $k$ is the the R-charge sector of the amplitude.

 The spacetime Lagrangian dictates that a tree level scattering amplitude in Einstein-Yang-Mills in the $\tau$ trace sector comes with a factor
\begin{equation}
 \kappa^{n_{gr} +2 \tau -2 }
 \end{equation} 
 of the gravitational coupling constant $\kappa \sim \sqrt {G_N}$, where $n_{gr}$ denotes the number of external gravitons. In \cite{Cachazo:2012kg} it was explained that, when written in terms of a worldsheet model, these powers of $\kappa$ must be accompanied by the same number of powers of $\langle \, , \rangle$ or $ [\, , ]$ brackets. Indeed, from dimensional analysis we find that 
 \begin{equation}
 \# \langle \, , \rangle + \# [\, ,] = n_{gr}^+ + n_{gr}^- +2 \tau -2 ~.
 \end{equation} 
 Parity conjugation exchanges $\langle \, , \rangle$ and $ [\, , ]$, which fixes
\begin{equation}
 \# \langle \, , \rangle  =   n_{gr}^- + \tau -1 ~,  \qquad   \# [ \, ,] = n_{gr}^+  +  \tau -1 ~.
\end{equation}
 From the perspective of twistor theory, the appearance of the $SL(2)_{L,R}$ invariants $\langle \, , \rangle$ and $[\, ,]$ controls the breaking of conformal symmetry of a theory, and the very existence of a well defined counting is a hallmark of the natural action (and breaking) of this symmetry on twistor space.

\subsection{Einstein-Yang-Mills amplitudes in 4d spinor helicity variables}  
 There are as many representations of any 4d refined scattering amplitude as there are representations of the 4d refined scattering equations themselves, and they each make different properties manifest. We begin with the non-supersymmetric ambitwistor representation, which makes parity manifest. Now the R-charge sector $k$ is simply given by the number of negative helicity particles, so $d = |\mathrm{p}^-|-1$. Using the known behaviour of the Jacobian~\cite{Cachazo:2013zc,Cachazo:2016sdc} which arises in going from the CHY to the refined scattering equations we find the Einstein-Yang-Mills scattering amplitudes in the $\tau$-trace sector
\begin{equation}\label{eqn:fullamplitudeAmbiTwistorspace}
\begin{aligned}
\int  \frac{1}{\text{vol GL}(2 , \mathbb{C}) }  &  \sum_{\substack{i_2 < j_2 \in \tr _2 \\  \cdots \\ i_\tau < j_\tau \in \tr _\tau }}     \det \left( \Phi ^{i \in \mathrm{h}^- \cup \mathrm{I}}  _{j \in \mathrm{h}^- \cup \mathrm{J}} \right)   ~ \det \left( \Phit    ^{i \in \mathrm{h}^+ \cup \mathrm{I}}  _{j \in \mathrm{h}^+ \cup \mathrm{J}}\right)  ~   \frac{ V( \mathrm{I} \cup \mathrm{J} ) }{   V(\mathrm{I}) ^2 V(\mathrm{J}) ^2 } \, \prod_{\alpha = 2 }^\tau  \, \frac{ ( z_{i_\alpha} - z_{j_\alpha}  ) }{  \diffd z_{i_\alpha}  \, \diffd z_{j_\alpha} }         \\
 &  \prod_{\alpha = 1}^\tau \text{PT}(\tr_\alpha) ~  \prod_{i \in \mathrm{p}^-} \frac{\diffd \tilde{u}_i}{\tilde{u}_i} \, \deltab^2( \lambdat_i - \tilde{u}_i \, \lambdat(z_i) )   \,  \prod_{i \in \mathrm{p}^+} \frac{\diffd u_i}{u_i} \, \deltab^2( \lambda_i - u _i \, \lambda(z_i) )  
\end{aligned}
\end{equation}
with the abbreviations
\begin{equation}
\mathrm{I} \equiv \{ i_2 , \cdots , i_\tau \}   \qquad \text{and} \qquad \mathrm{J} \equiv  \{ j_2 , \cdots , j_\tau \} ~.
\end{equation}
We use the familiar Hodges matrices (in the ambitwistor representation)
\begin{equation}
 \Phi_{ij} = \itw {\lambda_i } {\lambda_j} \, S(z_i,z_j) ~,\qquad  \Phi_{ii} = - \sum_{j \in \mathrm{p}^- \backslash \{i\}} \Phi_{ij} \, \frac{\tilde{u} _j}{\tilde{u} _i} 
\end{equation}
and
\begin{equation}
 \Phit_{ij} = \ditw {\lambdat_i } {\lambdat_j} \, S(z_i,z_j) ~,\qquad  \Phit_{ii} = - \sum_{j \in \mathrm{p}^+ \backslash \{i\}} \Phit_{ij} \, \frac{u _j}{u _i}  ~,
\end{equation}
for the integrand and the functions $\lambda(z) , \lambdat(z)$ are solutions to the ambitwistor scattering equations,
\begin{equation}
\lambda(z) = \sum_{i \in \mathrm{p}^-} \lambda_i \, \tilde{u}_i \, S(z,z_i) ~,\qquad \lambdat(z) = \sum_{i \in \mathrm{p}^+} \lambdat_i \, u_i \, S(z,z_i) ~,
\end{equation}
while $u_i , \tilde{u}_i$ are the corresponding scaling parameters. Furthermore we used the world-sheet Parke-Taylor factor of a gluon trace, defined as
\begin{equation}
\text{PT}(\tr) := \sum_{\sigma \in S_{|\tr|} / \mathbb{Z}_{|\tr|}}     \Tr \left[   \mathbf{T}_{\sigma(1)} \cdots   \mathbf{T}_{\sigma(|\tr|)} \right]       ~~   \prod_{i \in \tr} S(z_{\sigma(i)} , z_{\sigma( i+1)}) ~,
\end{equation}
with the gauge group generators $\mathbf{T}_i$ associated to each gluon in the trace. We emphasize again that while $\tr_1$ appears to be singled out, the scattering equations guarantee that the formula is independent of this choice, so is actually $S_\tau$ permutation symmetric. 

 It is well known that one of the major advantages of the refined formulas is that we can actually go beyond the bosonic amplitudes given by the CHY formula and find $\mathcal{N} \leq 3$ super-symmetric amplitudes in a remarkably simple fashion. Given the Grassmann numbers $\eta_i , \tilde{\eta}_{i}$ (transforming in the fundamental/anti-fundamental of the $SU(\mathcal{N})$ R-Symmetry, respectively) from the external supermomenta, we can promote \cref{eqn:fullamplitudeAmbiTwistorspace} to the full superamplitude by including the factor
\begin{equation}
\exp \left( \sum_{\substack{ i \in \mathrm{p}^- \\ i \in \mathrm{p}^+}} \eta_i \cdot \tilde{\eta}_j ~ \tilde{u}_i \, u_j  \, S(z_i,z_j) \right) ~,
\end{equation}
 whose behaviour under factorization is simple and well understood \cite{Adamo:2015gia}. This is astonishing not just because of its simplicity, but also because it makes space-time supersymmetry manifest. It is a consequence of the natural incorporation of onshell SUSY on twistor space.

\subsection{\Neqfour sEYM on Twistor Space} 
Now we shall give the twistor space representation of the scattering amplitude, which will break manifest parity invariance, but allow for manifest \Neqfour supersymmetry.

 In general, a scattering amplitude is a multi-linear functional of the external wave functions. Most commonly it is simply given in a basis of plane waves (as e.g. above), but on twistor space it is actually more natural to maintain the full structure. Using \Neqfour onhell SUSY we may write the wave function for a whole SUSY multiplet as a single function on onshell superspace. In \Neqfour sEYM there are two colour neutral multiplets, $h_i, \phi_i$, which contain the graviton as their highest/lowest spin state, and one adjoint-valued multiplet $ A _i$, containing the gluons.  Via the Penrose transform the external wave functions of the super-multiplets are given by cohomology classes with a certain homogeneity on super twistor space $\prtw := \prco^{3|4} \backslash \prco^{1|4}$
\begin{equation}
h_i \in H^1 \left( \prtw , \, \mathcal{O}(2) \right) ,~~  A_i \in H^1\left( \prtw , \, \mathcal{O}(0) \right) ,~~  \phi_i \in H^1 \left( \prtw , \, \mathcal{O}(-2) \right)  ,
\end{equation}
 of helicity $-2,-1,0$ respectively.  As usual, the coefficients in the Taylor expansion w.r.t. the Grassmann coordinates of $\prtw$ correspond to the various components of the supermultiplet. With these definitions in place we now present the sEYM scattering amplitude in the $\tau$ colour trace sector on twistor space
\begin{equation}\label{eqn:fullamplitudeTwistorspace}
\begin{aligned}
\sum_d \int _{\mathcal{M}_{0,n}(d)} \!\!\!\!\!\! \!\!\!\!\!\! \diffd \mu_d    \sum_{\substack{i_2 < j_2 \in \tr _2 \\  \cdots \\ i_m < j_m \in \tr _m }}     & \det \left( \Phi ^{i \in \mathrm{\phi} \cup \mathrm{I}}  _{j \in \mathrm{\phi} \cup \mathrm{J}} \right)   ~ \det \left( \Phit    ^{i \in \mathrm{h} \cup \mathrm{I}}  _{j \in \mathrm{h} \cup \mathrm{J}}\right)  ~   \frac{ V( \mathrm{I} \cup \mathrm{J} ) }{   V( \mathrm{I}) ^2 V(\mathrm{J} ) ^2 } \, \prod_{\alpha = 2 }^\tau  \, \frac{ ( z_{i_\alpha} - z_{j_\alpha}  ) }{  \diffd z_{i_\alpha}  \, \diffd z_{j_\alpha} }         \\
 &\prod_{\alpha = 1}^\tau \text{PT}(\tr_\alpha) ~  \prod_{i \in h}  h_i(Z(z_i))  ~ \prod_{i \in g}  A_i(Z(z_i))~ \prod_{i \in \phi} \phi_i(Z(z_i))
\end{aligned}
\end{equation}
where we abbreviated the sets
\begin{equation}
\mathrm{I} \equiv \{ i_2 , \cdots , i_\tau \}   \qquad \text{and} \qquad \mathrm{J} \equiv  \{ j_2 , \cdots , j_\tau \} ~,
\end{equation}
as well as the measure 
\begin{equation}
\diffd \mu _d \equiv \frac{ \diffd^{4(d+1)|4(d+1)} Z}{ \volGLtwo} \equiv \frac{\prod_{a=0}^d \diffd^{4|4} Z_a}{ \volGLtwo}
\end{equation}
 on $\mathcal{M}_{0,n}(d)$, the moduli space of holomorphic maps of degree $d$ from the $n$-punctured Riemann sphere to super twistor space. Here we have chosen to coordinatize this space as $Z(z) = \sum_{a=0}^d Z_a \, s_a(z)$, for some fixed basis\footnote{\label{ftn:measure}Note that since $\prtw$ is a Calabi-Yau supermanifold, the holomorphic measure $\diffd \mu_d$ is independent of the choice of basis $\{s_a\}_{a=0}^d$ by itself. Indeed, when the target space is $\prco^{m|\mathcal{N}}$, a change of basis in $H^0(\mathbb{P}^1 , \mathcal{O}(d))$ with Jacobian  $J(\{ s_a \}  ,  \{ s^\prime_a \})$  induces the integration measure to transform as $\diffd \mu _d \to \diffd \mu_d ~ J(\{ s_a \}  ,  \{ s^\prime_a \})^{m+1-\mathcal{N}}$.} of polynomials $\{s_a\}_{a=0}^d$ spanning $H^0(\mathbb{P}^1, \mathcal{O}(d))$. The Hodges matrices $\Phi , \Phit$ have been lifted to twistor space, with entries given by
 \begin{equation}
 \Phi_{ij} = S(z_i ,z_j)  \, \itw{Z(z _i) }{Z(z_j)}  ~, \qquad \Phi_{ii} =  \itw{Z (z_i) }{\diffd Z(z_i)}  
 \end{equation}
and
 \begin{equation}
 \Phit_{ij} = S(z_i ,z_j)  \,  \left[ \frac{\partial}{\partial Z(z _i)} \,  \frac{\partial}{\partial Z(z _j)} \right] ~, \qquad \Phit_{ii} = - \sum_{j \neq i} \Phit_{ij} ~ \frac{p(z_j)}{p(z_i)}   
 \end{equation}
 for some arbitrary section $p \in H^0(T^{1/2} \otimes \mathcal{O}(d))$. Here $\langle \cdot , \cdot \rangle $ and $[ \cdot , \cdot ]$ are the infinity twistor\footnote{\label{ftn:infinittTwistor}These are fixed simple bitvectors (antisymmetric matrices of rank 2) on twistor space, which arise in the decompactification of $\bar{ \mathbb{M}}$ to $\mathbb{M}$.} and dual infinity twistor\ftnref{ftn:infinittTwistor} respectively, i.e. $\langle Z Z^\prime \rangle = \mathcal{I}_{IJ} Z^I Z^{\prime J} $ for two twistors $Z , Z^\prime \in \prtw$ and $[ W W^\prime ] = \tilde{\mathcal{I}}^{IJ} W_I W^\prime_J $ for two dual twistors $W , W^\prime \in \prtw^\ast$. Generally, the appearance of the infinity twistor signals and controls the breaking of space-time conformal symmetry, which on twistor space is represented by general linear transformations. In the present case, they reduce as $\langle Z Z^\prime \rangle = \langle \lambda \, \lambda^\prime \rangle $ and $[ W W^\prime ] = [\lambdat \, \lambdat ^\prime ]$  to the Lorentz invariant pairings of left- and right-handed spinors, respectively.

\medskip
 There are two remarkable features of this formula. Firstly, on twistor space the dependence on the infinity twistor and dual infinity twistor has separated. This is akin to the separation in \Neqeight supergravity amplitudes \cite{Cachazo:2012kg,Skinner:2013xp}, but the addition of Yang-Mills interactions leads to a sum of such products. In other words, the presence of gluon traces obstructs a complete separation of the infinity twistor and dual infinity twistor, albeit in a rather systematic way. Secondly, the amplitude is now manifestly \Neqfour space-time supersymmetric, so \cref{eqn:fullamplitudeTwistorspace} includes amplitudes for space-time fermions. Neither of these properties is obvious/accessible from simply using the substitution $p_i \cdot p_i \to \itw{\lambda_i}{\lambda_j} \ditw{\lambda_i}{\lambda_j}$ in the dimension agnostic CHY formulae.

 Another property of the amplitudes can be learned from \cref{eqn:fullamplitudeTwistorspace}: since the Hodges determinant is an antisymmetric polynomial of degree $d-1$ in the marked points, it will vanish identically if $d < |\mathrm h | + \tau -1 $. Hence we find (for non-trivial amplitudes) the inequality $d +1 \geq  |\mathrm h | + \tau  $, and similarly the parity conjugate  $\dt +1 \geq  |\mathrm \phi | + \tau  $. Moreover, since the $R$-charge selection rules follow from the fermionic part of the map and wave-functions, which completely separates from the rest of the formula, we manifestly have the usual selection rules for \Neqfour SUSY (in particular $k = d+1$). This completely fixes the degree $d$ in terms of the external states, e.g. for external gravitons and gluons only, we recover $ d+1 = n_{gr}^- +  n_{gl}^- $ as expected. A corollary of this is that $n_{gl}^- \geq \tau $ and  $ n_{gl}^+ \geq \tau $, so any amplitude with less negative/positive gluons than traces will vanish.

 \medskip
 We may easily go from the twistor space representation back to the ambi-twistor representation \eqref{eqn:fullamplitudeAmbiTwistorspace} \cite{Geyer:2016nsh} by specifying the external states to be plane wave states and use the explicit form of the Penrose representative
 \begin{equation}
   \int \frac{\diffd t_i}{t_i } \, t_i ^{2s_i +2}    ~ \deltab^2(\lambda_i - t_i \, \lambda ) ~ \exp \left( t_i \, [\lambdat_i  \, \mu ] +  t_i \, \eta_i \cdot \chi \right)
 \end{equation} 
 for a multiplet of helicity $s_i$. Indeed, by judiciously choosing a coordinate basis\ftnref{ftn:measure} for the space of maps that is adapted to the external data,
 \begin{equation}
 Z(z) = \sum_{i\in \mathrm{p}^- } Z_i \, \prod_{j\in \mathrm{p}^- \backslash \{i\}} \frac{z-z_j}{z_i-z_j}
 \end{equation}
 while keeping the punctures fixed, we may perform the integral over the moduli space of the map $Z(z) $ trivially, upon which we recover the ambi-twistor representation.
 
It is also worth pointing out that \cref{eqn:fullamplitudeTwistorspace} reduces correctly to previously known $4d$ expressions. Indeed, in the pure gluon (single trace) sector it agrees immediately with the RSVW formula \Neqfour super Yang-Mills. In the pure graviton sector it reduces to the \Neqfour restriction of the CS formula, while the the enhancement to \Neqeight follows from the special properties of the Hodges determinants and scattering equations. A special case of this is given by the Einstein-Maxwell sector, where each trace contains exactly two gluons.

\section{Conclusion \& Outlook}

 We have presented and proven a formula for the splitting of certain fermion correlators into left and right handed Hodges type determinants. This factorization holds for general correlators of spinors with values in $T \mathbb{M}$ on the Riemann sphere, and in particular does not require the scattering equations to hold. We have then applied this splitting formula to the translation of d-dimensional CHY fomulas into the 4d spinor helicity formalism, which crucially involves the refined scattering equations, both for known examples and to find a new formula for all tree scattering amplitudes in \Neqfour EYM.

\bigskip
 The rational functions given in terms of Pfaffians and determinants that enter \cref{eqn:pfaffianFactorizationProto} have natural origins in $2d$ CFT on the punctured Riemann sphere. Even though we completely ignored this origin for the purpose of this paper, we believe that $2d$ CFT is the proper realm for understanding \cref{eqn:pfaffianFactorizationProto} (and in fact the crucial parts of the proof were discovered using that CFT description). Here we briefly sketch this relation, but leave the details for an upcoming publication.

 The left hand side of \cref{eqn:pfaffianFactorizationProto} is given by the correlator
 \begin{equation}
 \left \langle  \prod_{i=1}^{2n} (\lambda_i \lambdat_i) \cdot \psi(z_i) \right\rangle ~=~  \pf \bigg(  \itw{\lambda_i}{\lambda_j} \ditw {\lambdat_i}{\lambdat_j} \, S(z_i , z_j) \bigg)^{i,j=1,\cdots , 2n}
 \end{equation}
 with the action
 \begin{equation}
 S = \int_{\mathbb{P}^1} \psi^\mu \partialb \psi^\nu \, \eta_{\mu\nu} ~, \qquad\qquad \psi \in \Pi \Omega^0(K^{1/2} \otimes  T \mathbb{M}) ~,
 \end{equation}
 and no constraints on the locations $z_i$. On flat Minkowski space $ \mathbb{M} =\mathbb{R}^{3,1} $ the tangent bundle splits into a product of the left-handed and right-handed spin bundles
\begin{equation}
T \mathbb{M} ~\simeq~ \mathbb{S}^+ \otimes \mathbb{S}^- ~,
\end{equation}
 where the isomorphism is provided by the van der Waerden symbols $\sigma^\mu_{\alpha \alphad}$. We can write the world-sheet current\footnote{The argument in \cite{Zhang:2016rzb} is based on this identity.} associated to Lorentz transformations on either side of the isomorphism as
 \begin{equation}
 \psi^{[\mu}\psi^{\nu]} ~ \simeq~ \rho^a_\alpha \, \rho^b_\beta \, \varepsilon_{ab} ~ \varepsilon_{\alphad \betad} + \varepsilon_{\alpha \beta} ~   \rhot_{a \alphad}\, \rhot_{b\betad} \, \varepsilon^{ab} 
 \end{equation}
 with the new fields
\begin{equation}
\rho^a \in \Pi \Omega^0(K^{1/2} \otimes \mathbb{S}^- ) ~, \qquad  \rhot_a \in \Pi \Omega^0(K^{1/2} \otimes \mathbb{S}^+ ) ~,
\end{equation}
 where the Roman indices $a,b=1,2$ label the fundamental representation of a new $\text{SL}(2)$ gauge symmetry. This can be seen to arise as redundancy in the change of variables from $\psi $ to $\rho , \rhot$ and is responsible for the permutation symmetry of \eqref{eqn:buildingblock}. It is hence natural to suspect that the corresponding 2d sigma sigma models are also related. Indeed, the right hand side of \cref{eqn:pfaffianFactorizationProto} involves the correlators
\begin{equation}
\left \langle \prod_{i \in \mathrm{b}}  \langle \lambda_i  \rho ^1 (z_i)\rangle ~ \prod_{j \in \mathrm{b}^c}  \langle \lambda_j  \rho ^2 (z_j)\rangle  \right\rangle = \det \bigg(  \itw {\lambda_i}{\lambda_j} \, S(z_i,z_j) \bigg)^{i\in \mathrm{b}}_{j\in \mathrm{b}^c}
\end{equation}
in the action
\begin{equation}
S =\int_{\mathbb{P}^1} \varepsilon_{ab} \,  \langle \rho^a \partialb \rho ^b \rangle ~,
\end{equation}
and likewise for the right-handed determinant. Taking into account the appropriate Vandermonde determinants, we have proven (\cref{sec:proof}) that this correlator is indeed permutation symmetric. This suggests that there is an equivalence of CFTs that underlies the splitting formula \cref{eqn:pfaffianFactorizationProto}.

Closely related to the previous comments is the fact that the twistor space connected formula \eqref{eqn:fullamplitudeTwistorspace} for EYM amplitudes can naturally be incorporated into the twistor string of \cite{Skinner:2013xp}, which we describe in an upcoming publication.

\medskip
Finally, it is very tempting to apply \cref{eqn:pfaffianFactorizationProto} to scattering-equation based formulas for higher-loop amplitudes, which currently come in two flavours. On the one hand, the ambitwistor string model \cite{Mason:2013sva} gives rise to amplitudes for supergravity on the torus \cite{Adamo:2013tsa}. Indeed, \cref{eqn:pfaffianFactorizationProto} has a natural generalization to higher genus surfaces, and we expect a generalization of the proof here to carry over. It is however believed that the ambitwistor string is only modular invariant in $10d$, so even though the external states can easily be restricted to lie in a 4d subspace, the loop momentum would have to be integrated over a $10d$ space, which will make $P(z)$ generically a $10d$ vector. This obstructs the use of \cref{eqn:pfaffianFactorizationProto} as shown here, since the $C_{ii}$ elements of the CHY type Pfaffian cannot be split straightforwardly. Further complications might arise from Ramond sector fields or the spin-structure dependence of the Szeg\'o kernel.

On the other hand there are formulas for loop amplitudes on the nodal sphere \cite{Geyer:2015bja,Geyer:2015jch,Geyer:2016wjx}. While these seem to be well defined (or at least come with a canonical regularization scheme) in any dimension, the above obstruction remains: on the nodal sphere the scattering equations imply generically $P(z)^2 \neq 0$, so we again cannot split the $C_{ii}$ elements of the CHY type Pfaffian straightforwardly. In this situation the resolution might be more apparent: we can write $P(z)$ as a sum of null vectors, and, using the multi-linearity of the Pfaffian, apply the factorization \cref{eqn:pfaffianFactorizationProto} to each summand separately. There have just been promising new results \cite{Farrow:2017eol} for $4d$ loop amplitudes based on the $4d$ refinement of the scattering equations on the nodal sphere, which might be combined naturally with the present work to find $n$-point SUGRA integrands. We leave these exciting thoughts and questions for future work.

\section*{Acknowledgements}
 It is a privilege to thank David Skinner for many inspiring and insightful discussions as well as continuous encouragement. The author would also like to thank Tim Adamo, Piotr Tourkine and Lionel Mason for interesting discussions.

The author is supported in part by a Marie Curie Career Integration Grant (FP/2007-2013/631289).

\appendix 

\section{Proof of Chiral Splitting in 4d}  \label{sec:proof}
 In this appendix we give the full details of the proof of \cref{eqn:pfaffianFactorizationProto}, which we repeat here for the reader's convenience
 \begin{equation}
 \pf \left( \frac{\itw ij \ditw ij}{z_i - z_j} \right)^{i,j=1,\cdots , 2n} =  ~\frac{\det \left( \frac{\itw ij}{z_i - z_j} \right)^{i\in \mathrm{b}}_{j\in \mathfrak{b}^c} }{V( \mathrm{b}) \, V( \mathrm{b} ^c)}   ~~ \frac{\det \left( \frac{\ditw ij}{z_i - z_j} \right)^{i\in \tilde{ \mathrm{b}}}_{j\in \tilde{ \mathrm{b}}^c} }{V( \tilde{ \mathrm{b}}) \, V( \tilde{ \mathrm{b}} ^c)} ~~ V( \{ 1 , \cdots , 2n \} )
 \end{equation}
 where $ \mathrm{b}, \tilde{ \mathrm{b}}$ are any ordered subsets of $\{1, \cdots , 2n \}$ of size $n$ and $\tilde{ \mathrm{b}} ^c, \mathrm{b}^c$ are their complements. First we recall the definitions. Take $2n$ points on the Riemann sphere, given in inhomogeneous coordinates by $z_i$, as well as one left- and right-handed spinor $\lambda_{i}, \lambdat_i$ associated to each puncture, with $i = 1 ,\cdots , 2n$. We also have the Vandermonde determinant of an ordered set, defined as usual
 \begin{equation}
 V( \mathrm{b} ) := \prod_{i<j \in \mathrm{b}} (z_i - z_j) 
 \end{equation}
 We begin by proving that the building block
\begin{equation}
\frac{\, \det \left( \frac{\itw ij}{z_i - z_j} \right)^{i\in \mathrm{b}}_{j\in \mathrm{b}^c} }{V( \mathrm{b}) \, V( \mathrm{b} ^c)}  
\end{equation}
is independent of the split of the labels $\{1 , \cdots , 2n \}$ into the two ordered subsets $\mathrm{b} , \mathrm{b}^c$ -- in other words, it is still $S_{2n}$ permutation symmetric, albeit not manifestly so.

\subsection{Proof of $S_{2n}$ Symmetry}
 For definiteness, and without loss of generality, we assign the labels $\{1 , \cdots , n \}$ to the rows and $\{n+1 , \cdots , 2 n \}$ to the columns, so we prove the $S_{2n} $ permutation symmetry of
\begin{equation}\label{eqn:AGeneralizedHodgesMatix}
 ~\frac{ \det \left( \Phi ^{i = 1 , \cdots , n} _{j = n+1 , \cdots , 2n} \right)  }{V(\{ 1, \cdots ,n\} ) ~ V(\{n+1, \cdots ,2n\} )}
\end{equation}
 which is used in the main text. Recall that the numerator is given by the determinant of the $n\times n$ matrix $\Phi$ with elements
 \begin{equation}
 \Phi^i_j = \frac{ \langle \lambda_i , \lambda_j \rangle }{z_i-z_j} ~,\qquad \text{for } i = 1, \cdots ,n ~\text{ and } ~ j = n+1 , \cdots , 2n  ~,
 \end{equation}
 where $\langle \cdot , \cdot \rangle $ is the $SL_2$ invariant pairing of the left-handed spinors. We emphasize that while the set of punctures that label the rows is disjoint from the set of punctures that label the columns, we can easily achieve any overlap between the sets labelling rows and columns by taking an appropriate limit of the above matrix, with a ``diagonal element'' that we can specify freely. This is necessary e.g. for amplitudes involving gravitons.
 
 Notice that, due to the antisymmetry of the Vandermonde determinant as well as the numerator \cref{eqn:AGeneralizedHodgesMatix} is manifestly $S_n \times S_n \times \mathbb{Z}_2$ symmetric, i.e. permutation symmetric in each of the two sets $\{1 , \cdots , n \}$ and $\{n+1 , \cdots , 2n \}$ separately, while the $\mathbb{Z}_2$ factor swaps the two sets and transposes the matrix. In order to show that it has in fact $S_{2n} $ symmetry we will show that as a rational function of the locations $z_i$ and spinors $\lambda_i$ it is equal to the expression 
\begin{equation}\label{eqn:AGeneralizedHodgesMatix2}
   (-1)^{\frac{n(n-1)}{2}} \sum_{\substack{ \mathrm{b} \subset \{ 1, \cdots , 2n\} \\ |\mathrm{b}| =  n  }}  ~ \prod_{i \in \mathrm{b}} (\lambda_i)^0  ~ \prod_{j \in \mathrm{b}^c} (\lambda_j)^1 \prod_{ \substack{ i \in \mathrm{b}  \\ j \in \mathrm{b}^c}} \frac{1}{z_i - z_j}
\end{equation}
 Since this expression is manifestly $S_{2n}$ invariant (even though it has lost its manifest  Lorentz $\text{SL}_2$ invariance), proving equality of \cref{eqn:AGeneralizedHodgesMatix} and \cref{eqn:AGeneralizedHodgesMatix2} will establish the $S_{2n}$ symmetry of \cref{eqn:AGeneralizedHodgesMatix}.

The plan is to examine the poles and residues of each expression as any of the two punctures coincide, and then use a recursion argument to show that the residues agree. First, we rewrite the claim as
\begin{equation}
  \det \left( \Phi ^{i = 1 , \cdots , n} _{j = n+1 , \cdots , 2n} \right)    = (-1)^{\frac{n(n-1)}{2}} \!\!\!\!\!\!   \sum_{\substack{ \mathrm{b} \subset \{ 1, \cdots , 2n\} \\ |\mathrm{b}| =  n  }}  ~ \prod_{i \in \mathrm{b}} (\lambda_i)^0  ~ \prod_{j \in \mathrm{b}^c} (\lambda_j)^1  ~ \frac{V(1, \cdots ,n ) \, V(n \!+ \! 1, \cdots ,2n )}{   \prod_{ \substack{ i \in \mathrm{b}  \\ j \in \mathrm{b}^c}f} (z_i - z_j)} ~,
\end{equation}
 Each side is now a section of $\otimes _i \mathcal{O}_i(-1)$ with at most simple poles as any two punctures coincide, so by Cauchy's theorem, comparing residues is sufficient to prove equality. Furthermore, given the already manifest $S_n \times S_n \times \mathbb{Z}_2$ symmetry, it is sufficient to check the residues at $ z_1= z_2 $ and  $ z_1 = z_{2n} $. It is actually immediately clear that both sides have vanishing residue at $ z_1 = z_2 $, so we only have to put some effort into checking the residue at $z_1 =  z_{2n} $. On the left hand side we find
 \begin{equation}
 \lim _{ z_1 \to z_{2n} } \left\{  (z_1 -  z_{2n}) ~ \det \left( \Phi ^{i = 1 , \cdots , n} _{j = n+1 , \cdots , 2n} \right)   \right\} = (-1)^n\,  \langle \lambda_1 , \lambda_{2n} \rangle \, \det \left( \Phi ^{i = 2 , \cdots , n} _{j = n+1 , \cdots , 2n-1} \right)  ~,
 \end{equation}
while on the right hand side we find
 \begin{equation}
 \begin{aligned}
 \lim _{ z_1 \to z_{2n} } & \left\{  (z_1 -  z_{2n})  \sum_{\substack{ \mathrm{b} \subset \{ 1, \cdots , 2n\} \\ |\mathrm{b}| =  n  }}  ~ \prod_{i \in \mathrm{b}} (\lambda_i)^0  ~ \prod_{j \in \mathrm{b}^c} (\lambda_j)^1  ~ \frac{V(1, \cdots ,n ) ~ V(n+1, \cdots ,2n )}{   \prod_{ \substack{ i \in \mathrm{b}  \\ j \in \mathrm{b}^c}}     (z_i - z_j) }    \right\}  \\
  & =  \langle \lambda_{2n} ,\lambda_1  \rangle \,  \sum_{\substack{ \mathrm{b} \subset \{ 2, \cdots , 2n-1\} \\ |\mathrm{b}| =  n-1  }}  ~ \prod_{i \in \mathrm{b}} (\lambda_i)^0  ~ \prod_{j \in \mathrm{b}^c} (\lambda_j)^1  ~ \frac{V(2, \cdots ,n ) ~ V(n+1, \cdots ,2n -1)}{   \prod_{ \substack{ i \in \mathrm{b}  \\ j \in \mathrm{b}^c}}  (z_i - z_j) }
 \end{aligned}
 \end{equation}
 where, going to the second line, we observed that only those terms in the sum where $1 $ and $2n$ are in different subsets contribute to the pole. We immediately recognize the condition for the residues to agree as the very same claim we're trying to prove but for $n-1$. Hence, we may conclude the proof by invoking a simple induction argument from $n$ to $n-1$.
 

 \subsection{Proof of Splitting}
 Armed with the knowledge that the factor \cref{eqn:AGeneralizedHodgesMatix2} is secretly $S_{2n}$ symmetric, we may now establish the factorization formula \cref{eqn:pfaffianFactorizationProto} by comparing residues. Both sides are again sections of $\otimes_i \mathcal{O}_i(-1)$ so comparing residues as any pair of punctures collide is sufficient to prove equality.
 
 Using the $S_{2n}$ symmetry of both sides we may simply look at the residue as $z_1 \to z_2$, where we find for the left hand side
 \begin{equation}
  \lim _{ z_1 \to z_2 } \left\{  (z_1 -  z_2) \,  \pf \left( \frac{\itw ij \ditw ij}{z_i - z_j} \right)^{i,j=1,\cdots , 2n} \right\} =  \itw 12 \ditw 12 ~  \pf \left( \frac{\itw ij \ditw ij}{z_i - z_j} \right)^{i,j=3,\cdots , 2n} ~.
 \end{equation}
 On the right hand side we first make a judicious choice for the splitting of labels into rows and columns such that the rows of the first matrix be labelled by the set $\{1\} \cup \mathrm{b}^\prime $ and the columns by $\{2\} \cup {\mathrm{b}^\prime}^c $ where ${\mathrm{b}^\prime} \cup {\mathrm{b}^\prime}^c  = \{3 , \cdots , 2n \}$ is a partition of the remaining labels and similarly for the second matrix. (For the sake of clarity we drop the primes below.) Hence we find the residue on the right hand side
 \begin{equation}
 \begin{aligned}
 \lim _{ z_1 \to z_2 } & \left\{  (z_1 -  z_2)  ~  \frac{\det \left( \frac{\itw ij}{z_i - z_j} \right)^{i\in\{1\} \cup \mathrm{b}}_{j\in \{2\} \cup   \mathrm{b}^c } }{V( \{1\} \cup \mathrm{b}) \, V(\{2\} \cup \mathrm{b}^c )}   ~~ \frac{\det \left( \frac{\ditw ij}{z_i - z_j} \right)^{i\in  \{1\} \cup  \tilde{ \mathrm{b}}  }_{j\in  \{2\} \cup  \tilde{ \mathrm{b}} ^c } }{V(\{1\} \cup  \tilde{  \mathrm{b}}) \, V( \{2\} \cup  \tilde{ \mathrm{b}} ^c)} ~~ V( \{ 1 , \cdots , 2n \} )  \right\} \\
 & {}=  \itw 12 \ditw 12 ~  \frac{\det \left( \frac{\itw ij}{z_i - z_j} \right)^{i\in  \mathrm{b}}_{j\in   \mathrm{b}^c } }{V(  \mathrm{b}) \, V(  \mathrm{b}^c )}   ~~ \frac{\det \left( \frac{\ditw ij}{z_i - z_j} \right)^{i\in   \tilde{ \mathrm{b}}  }_{j\in    \tilde{ \mathrm{b}} ^c } }{V(  \tilde{  \mathrm{b}}) \, V(   \tilde{ \mathrm{b}} ^c)} ~~ V( \{ 3 , \cdots , 2n \} ) 
 \end{aligned}
 \end{equation}
 Notice that each determinant has a simple pole as $z_1 \to z_2$, while the big Vandermonde factor has a simple zero, and on the location of the residue there are cancellations between the various Vandermonde factors.  We again recognize the condition for the residues to agree as the very same claim we're trying to prove but for $n-1$ so, after invoking recursion, this concludes the proof of the splitting formula \cref{eqn:pfaffianFactorizationProto}.


\bibliography{main}
\bibliographystyle{JHEP}

\end{document}